\documentclass[sigconf]{acmart}
\renewcommand\footnotetextcopyrightpermission[1]{} % removes footnote with conference information in first column

\usepackage[flushleft]{threeparttable}
\usepackage{epsfig}
\usepackage{endnotes}
\usepackage{setspace}
\usepackage{array}
\usepackage{booktabs}
\usepackage{graphicx}
\usepackage{amsfonts}           
\usepackage{mathrsfs}
\usepackage[ruled,vlined]{algorithm2e}
\usepackage{comment}
\usepackage{xargs}
\usepackage{mathtools}
\usepackage{balance}
\usepackage{tabularx}
\usepackage{subfigure}
\usepackage{enumerate}
\usepackage{amsthm}
\usepackage{enumitem}
\usepackage{xspace}
\usepackage{tikz}
\usepackage{multirow}
\usepackage{lipsum}
\usepackage[htt]{hyphenat}
\usepackage{listings}
\usepackage{color}
\usepackage{xcolor}
\usepackage{graphicx}
\usepackage{lscape}
\usepackage{algorithmic}
\usepackage{pifont}
\usepackage{multirow}
\usepackage{blindtext}
\usepackage[framemethod=TikZ]{mdframed}

\acmConference{}
\acmYear{}

\def\etc{\emph{etc}\xspace}

\definecolor{dkgreen}{rgb}{0,0.6,0}
\definecolor{gray}{rgb}{0.5,0.5,0.5}
\definecolor{mauve}{rgb}{0.58,0,0.82}
\definecolor{applegreen}{rgb}{0.55, 0.71, 0.0}
\definecolor{amber}{rgb}{1.0, 0.75, 0.0}
\definecolor{firebrick}{rgb}{0.7, 0.13, 0.13}
\definecolor{darkblue}{rgb}{0,0,0.55}

\begin{document}

\title{Visualization and Attack Prevention for a Sensor-Based Agricultural Monitoring System}

\author{Yifan Zhou$^{\ast}$}
\affiliation{%
  \institution{The University of Adelaide}
  \country{Australia}}
\author{Zhendong Shi$^{\ast}$}
\affiliation{%
  \institution{The University of Adelaide}
  \country{Australia}}
\author{Ruoxi Sun}
\affiliation{%
  \institution{The University of Adelaide}
  \country{Australia}}

\renewcommand{\shortauthors}{Y. Zhou, Z. Shi, and R. Sun}
 \renewcommand \authors{Yifan Zhou, Zhengdong Shi, and Ruoxi Sun}

\begin{abstract}
This project proposes a sensor-based visual agricultural monitoring system. Distinguished from traditional agricultural monitoring systems, this system further analyzes basic agricultural data and prevents and monitors common wireless network attacks such as Selective Forwarding, Black Hole Attacks, Sinkhole Attacks, Flooding Attacks and Misdirection Attacks. Experimental verification and evaluation of the attack prevention and monitoring are also conducted.{\let\thefootnote\relax\footnote{{$\ast$~The two authors contributed equally.}}}

\end{abstract}

\maketitle

\section{Introduction}
With the development of IoT technology, more and more wireless sensor networks are employed in diverse fields, such as agriculture, industry, and medical monitoring, to transmit the data collected by sensors~\cite{prasad2020IoT}.

Precision agriculture is one of the most popular filed. In precision agriculture, environmental data such as temperature and humidity collected by wireless sensors are monitored and analyzed to help farmers make accurate judgments~\cite{jana2019analysis}.

% In this way to help farmers manage their crops and provide intelligent advice. 
% It combines technology and management to improve agricultural production efficiency.

% Web-based wireless sensor network monitoring systems are widely used in agriculture~\cite{stafford2000implementing}, industry and medical fields. Whether it is applied to the collection and analysis of agricultural data or the monitoring of medical information. 

With the rapid growth of using mobile applications and IoT in modern daily life, security challenges in authentication, data storage, management, and access control threaten the security and privacy of users more and more~\cite{Feng2021SnipuzzBF,sun2021empirical,sun2020quality,sun2020venuetrace}. For example, attacking agricultural sensors can lead to biased data analysis results of the system and eventually lead to damaged agricultural products. Attacking the IoT infrastructure can lead to the power system going down~\cite{khanam2020survey}.
Although wireless sensor network monitoring systems are so widely utilized by various manufacturers, there are still many designers who ignore the security issues involved~\cite{akyildiz2002wireless}.
The security of wireless sensor networks is often overlooked. Whether it is in agriculture, industry, medical monitoring or smart home, many sensors are vulnerable to attacks, and controllers such as microcontrollers that control the sensors are also vulnerable to attacks. Some wireless transmission modules and hardware have very low security when transmitting data, and some even use broadcast transmission, which is very detrimental to data security. In this case, it is easy for attackers to steal data or use large amounts of invalid data to crash the device. 
\begin{figure*}[t]
	\centering
	\includegraphics[width=0.9\linewidth]{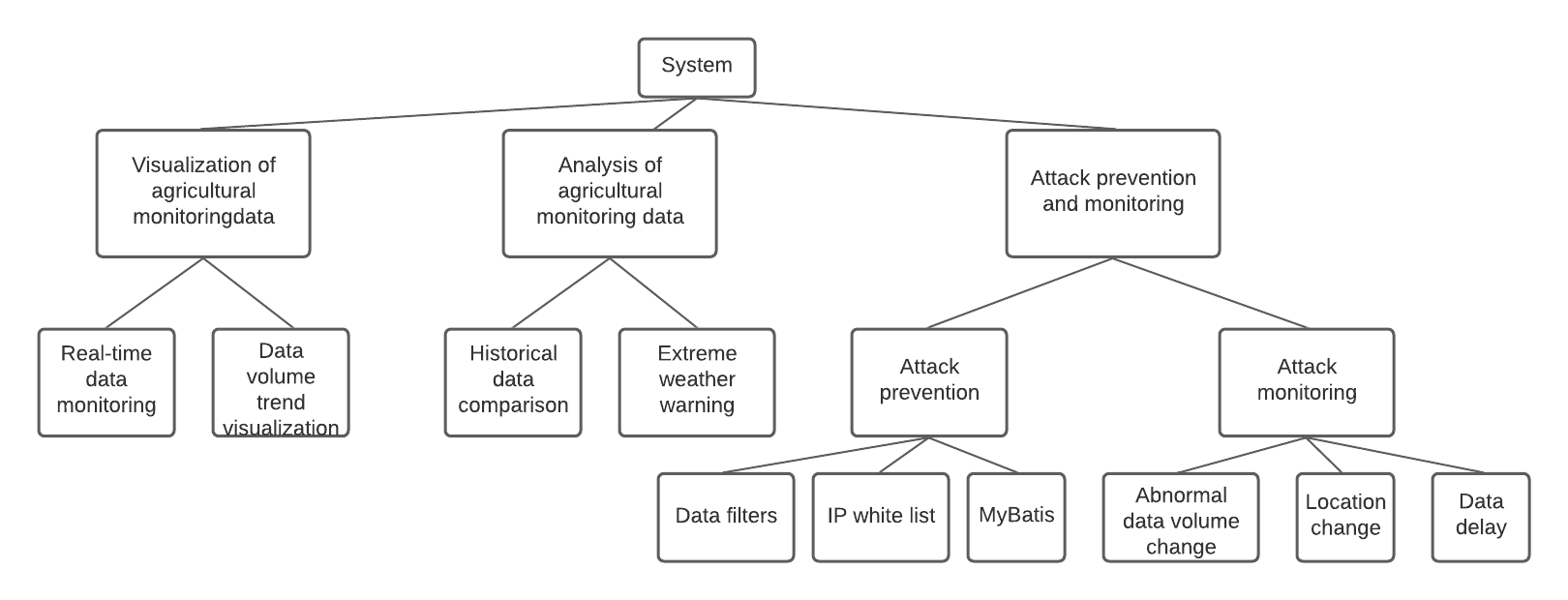}.
	\vspace{-3mm}
	\caption{System design.}
	\label{fig:figoverview}
\end{figure*}

\begin{figure*}[t]
	\centering
	\includegraphics[width=0.9\linewidth]{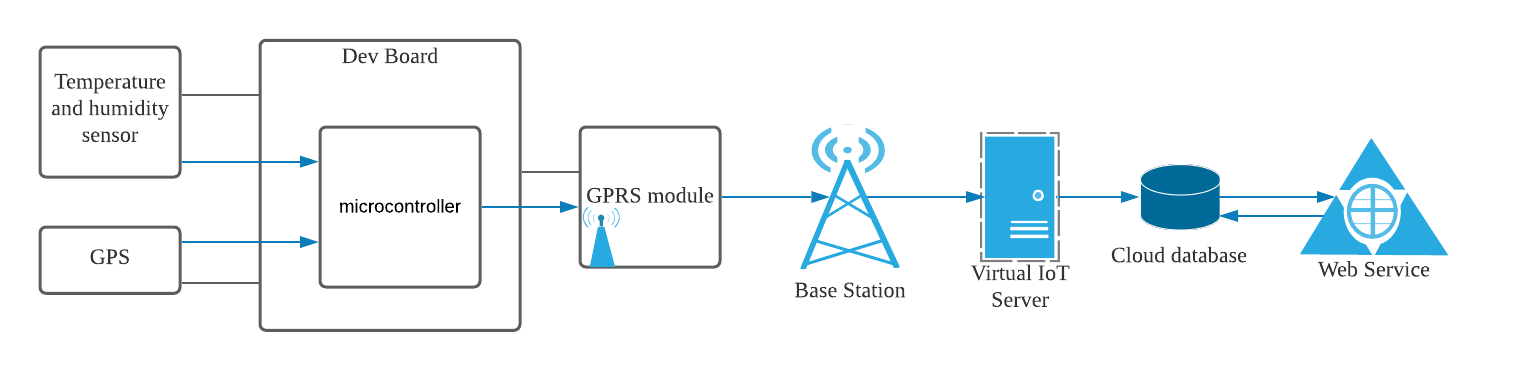}.
	\vspace{-3mm}
	\caption{Architecture of the system.}
	\label{fig:figp}
\end{figure*}

In this research, we propose a visualization and attack prevention approach to establish a sensor-based monitoring system, enabling data monitoring and analysis, as well as attack monitoring and prevention in agricultural management. The proposed system can effectively defense against wireless network attacks, such as Jamming, Tampering, and Exhaustion~\cite{sunilkumar2017review,hari2016security}, by analyzing the characteristics of attacks.

Different from the existing smart agricultural management systems~\cite{kubicek2013prototyping,haswani2018web}, our system not only implemented the visual monitoring of agricultural data and provided basic data query and comparison functions, but also further enabled the data analysis which can provide helpful recommendations automatically. Furthermore, with data transmission security in our mind, we proposed several defenses and monitoring approaches against potential attacks. The key contributions of this research are as follows.

\begin{itemize}
    \item We proposed an approach to enhance the precious agricultural with visualization and analysis of agricultural monitoring data and prevention and monitoring of attacks.
    \item We implemented a system that visualizes monitoring agricultural, provides basic data query and comparison functions, performs further analysis of data, prevention and monitoring of attacks.
    \item We evaluate the system against selective forwarding, black hole attacks, sinkhole attacks, flooding attacks, misdirection attacks and physical attacks. We have also conducted some experiments to verify these attack monitoring and prevention. The experimental results indicate that our system is well protected against these attacks and has a good performance in the timeliness of attack warnings.
\end{itemize}

% Many existing systems have implemented visual monitoring of agricultural data and provided basic data query and comparison functions~\cite{kubicek2013prototyping,haswani2018web}, but few of them perform further analysis of the data, making recommendations based on the data that will be helpful to agriculture\cite{karimi2018web}. Even fewer systems take into account the security of data transmission and do not prevent and monitor attacks. Our system not only has the functions of a traditional agricultural monitoring system, but also performs further analysis of data, prevention and monitoring of attacks.

\section{Background and Related Work}

\vspace{1mm}
\noindent \textbf{Wireless sensor networks.~} Wireless sensor networks have widely used in our life, such as smart home, security system, \etc. By introducing Internet access capabilities, wireless sensor networks and IoT also have collaboration. Such collaboration also brings new challenges in security~\cite{butun2019security}.

A common wireless sensor network based system goes through many steps from collecting data to transmitting data to the cloud, and each step can be attacked, which can lead to data leakage, loss, device crash, \etc. There are many ways to conduct attacks, the most common ways are forwarding data, dropping data, creating fake data, and transmitting a large number of empty packets. Therefore, the security of wireless sensor networks has become an issue that needs to be considered seriously.

\vspace{1mm}
\noindent \textbf{IoT application in precious agricultural.~} The collaboration of wire sensor networks and the Internet of Things also has a wide range of applications in the field of precision agriculture. Specifically, wireless sensor networks benefit the monitoring of pesticides level in the drinking water, the level of soil erosion, and the level of air pollution in real time~\cite{avinash2020wireless}. It uses sensors to collect data, such as temperature, humidity, and nutrient content in the soil~\cite{rajendra2020IoT}. The sensor data is further analyzed and suggestions are given to maintain the crop growth environment automatically. Such a smart agriculture solution enhances modern agriculture management with state-of-the-art information technology, improving agricultural production efficiency significantly.

\section{System Design}
In this section, we introduce the design of a sensor-based agricultural monitoring system that visualizes the agricultural data and monitors potential attacks at the same time. As shown in Figure~\ref{fig:figoverview}, the system consists of three main components: the visualization of agricultural monitoring data, the analysis of agricultural monitoring data, the sensor attack prevention and monitoring.

\subsection{Architecture overview}
As shown in Figure~\ref{fig:figp}, temperature and humidity data are captured by sensors and location information is obtained by the GPS module. The microcontroller processes the acquired data and uploads it to the base station through the GPRS module. The data is then transferred to a virtual IoT server for filtering and processing. The final processed data is stored in the cloud database. The web service will interact with the cloud database to implement the system's functions by querying and analyzing the data.

\subsection{Visualization of agricultural monitoring data}
\begin{figure*}[t]
	\centering
	\includegraphics[width=0.95\linewidth]{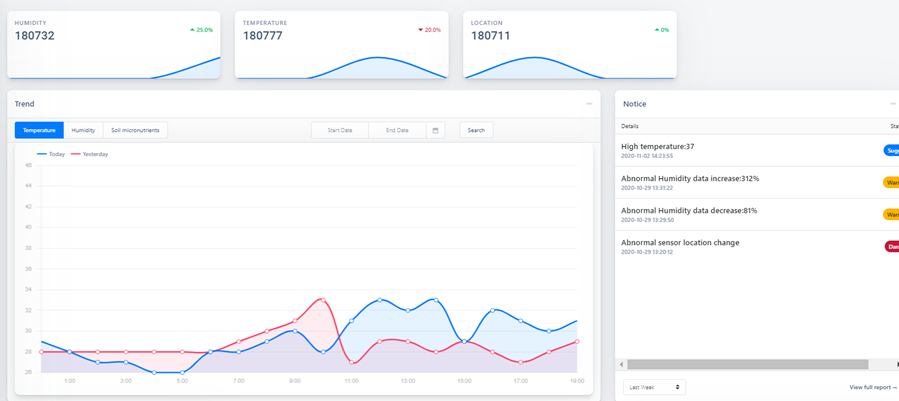}
	\vspace{-3mm}
	\caption{Visualization and analysis of agricultural monitoring data.}
	\label{fig:figmonitor}
\end{figure*}

Visualization of agricultural monitoring data is to transmit the data collected by sensors to the control center through wireless, then send them to the cloud database after filtering, and finally send them to the web page to show in the form of graphs.

This system needs to be implemented to collect the data from the agricultural sensors via wireless and collate the data and upload it to the cloud server at regular intervals. Based on these requirements, we chose the model STM32F407 development board as the core of the hardware part, it comes with a high-performance microcontroller that supports RS485 format transmission, it is a good choice to process the data collected from the sensors.

Because the most important and basic data in agricultural monitoring is temperature and humidity, and in this system, we want to implement the function of reminding farmers to water according to the change of data, so we chose the temperature and humidity sensors. In order to achieve accurate real-time positioning, we connected an external GPS module on the development board.
	
The board transmits the data through the GPRS module and the data is transferred to the base station using the MQTT protocol. Then to the IoT virtual server. The data will be forwarded by the IoT platform to the cloud database.
	
The web page will send requests to the back end at regular intervals to query the latest data in the database and analyze the data. If the analysis result determines that there is a suspicious attack or farmers need to water, it will return the latest data and warning message or suggestions to the web front end and store that in the database.	

The basic data our agricultural monitoring system mainly shows is the total amount of temperature and humidity data collected, and displays it in the form of a graph. The numbers in the graph indicate the total amount of accumulated data, and show the trend of increase and decrease of data volume for the last 160s. It can be clearly seen whether there is any abnormal change in the data volume. The rate of change therein indicates the relative increase or decrease of data volume for the last 40s and 40s-80s. If there is a significant increase or decrease in the percentage value of the rate of change, an attack may be present.

\begin{figure*}[t]
	\centering
	\includegraphics[width=0.98\linewidth]{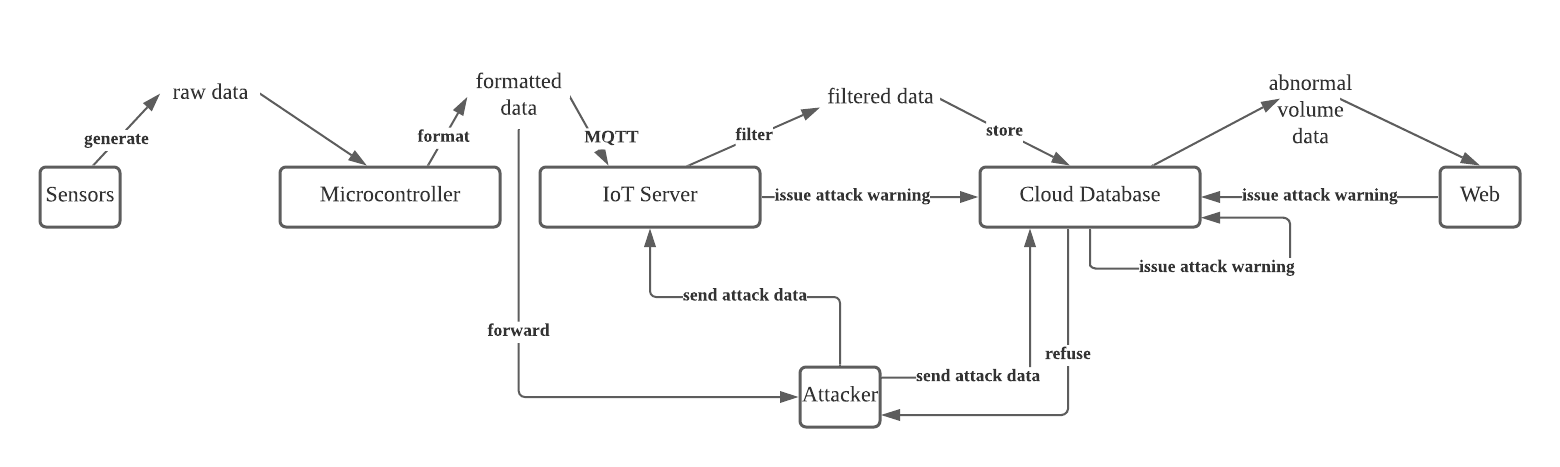}.
	\vspace{-3mm}
	\caption{Attack prevention and monitoring.}
	\label{fig:figattack}
\end{figure*}

\subsection{Analysis of agricultural monitoring data}

Further analysis of the agricultural monitoring data is to analyze and compare the data. The real-time data is compared with the historical data, and the comparison results are displayed in the form of graphs and charts, and some agricultural suggestions are given to farmers according to the changes in the data values, such as the need for watering, fertilizing, \etc. It also warns about the anomalies of the data.

The historical data shows temperature and humidity data for 24 hours by default, and is displayed as a line graph. When a specified hour is selected, the data for that period is displayed, along with comparative data for the same period of the previous day.

It is also possible to select a specific period to query through the query function. The system supports data queries within one year, with each query spanning a week. Also, data categories (such as temperature or humidity) can be filtered. If the selected span is one week, then the unit will become days.

The graph will show the highest and lowest values for each day of the selected data type for that period for comparison. This allows us to clearly see how the data values have changed over the selected period. From this, we can analyze whether too much water has evaporated or whether the crops need to be kept warm by cooling down too quickly.

If the air temperature is higher than the preset maximum value or lower than the preset minimum value for a certain period, an extreme temperature warning will be issued. If the air humidity is below or above a certain value, a warning will also be appeared in the warning table.
% \begin{figure*}[t]
%   \centering
%     \subfigure[]{\includegraphics[width=0.8\linewidth]{figures/data analysis.png}}
%     \subfigure[]{\includegraphics[width=0.5\linewidth]{figures/notice.png}}
%   \caption{(a) Analysis of historical data; (b) Warning table.}
%     \label{fig:data_analysis}
%     \vspace{0.2In}
% \end{figure*}

\subsection{Attack prevention and monitoring}
Sensor attack prevention and sensor attack monitoring are to detect and monitor attacks by analyzing the data and comparing the characteristics of various attack behaviors. For example, for attacks that transmit a large amount of useless data, these data will be filtered and an alert message will be sent to the front end to warn the user.

% \section{Attack prevention and monitoring}
\subsubsection{Common types of wireless network attacks}
In this research, we considered several state-of-the-art attacks against the wireless sensor networks, such as Selective Forwarding, Black Hole Attacks, Sinkhole Attacks, Flooding Attacks, and Misdirection Attacks~\cite{chhaya2017wireless,butun2019security,wang2006survey,cayirci2008security}.

\vspace{1mm}
\noindent \textbf{Black Hole Attacks~\cite{butun2019security}.~} The attacker forwards the data through the malicious node, causing the data traffic to stop, just like throwing them into a black hole. The data volume drops rapidly in a short period.

\vspace{1mm}
\noindent \textbf{Selective Forwarding~\cite{chhaya2017wireless,butun2019security,karlof2003secure}.~} The attacker forges a node, forwards data collected by the sensor to his/her designated address to steal data, and drops some packets. Sometimes this attack will behave like a black hole attack, dropping all packets.

\vspace{1mm}
\noindent \textbf{Sinkhole Attacks~\cite{karlof2003secure}.~} The attacker misleads the routing of nodes to attract them\cite{karlof2003secure}. It can attract nearby traffic through a controlled node. This gives access to the packets transmitted by the nodes, followed by other attacks, such as performing Selective Forwarding or Black Hole Attacks.

\vspace{1mm}
\noindent \textbf{Flooding Attacks~\cite{butun2019security}.~} The attacker will maliciously send broadcasts to the node, causing malicious packets to occupy the channel, which leads to the loss of information transmitted by itself, or even causes the node to crash due to resource exhaustion.

\vspace{1mm}
\noindent \textbf{Misdirection Attacks~\cite{chhaya2017wireless,butun2019security}.~} Data is routed to the wrong path, which results in the target node not receiving any packets, which leads to loss or delay of data.

The impact of these attacks on this system will be mainly in four aspects: data loss, a big amount of fake data, data delay, and node crash. Therefore, we will focus on monitoring and preventing these phenomena.

\begin{figure*}[t]
  \centering
    \subfigure[]{\includegraphics[width=0.48\linewidth]{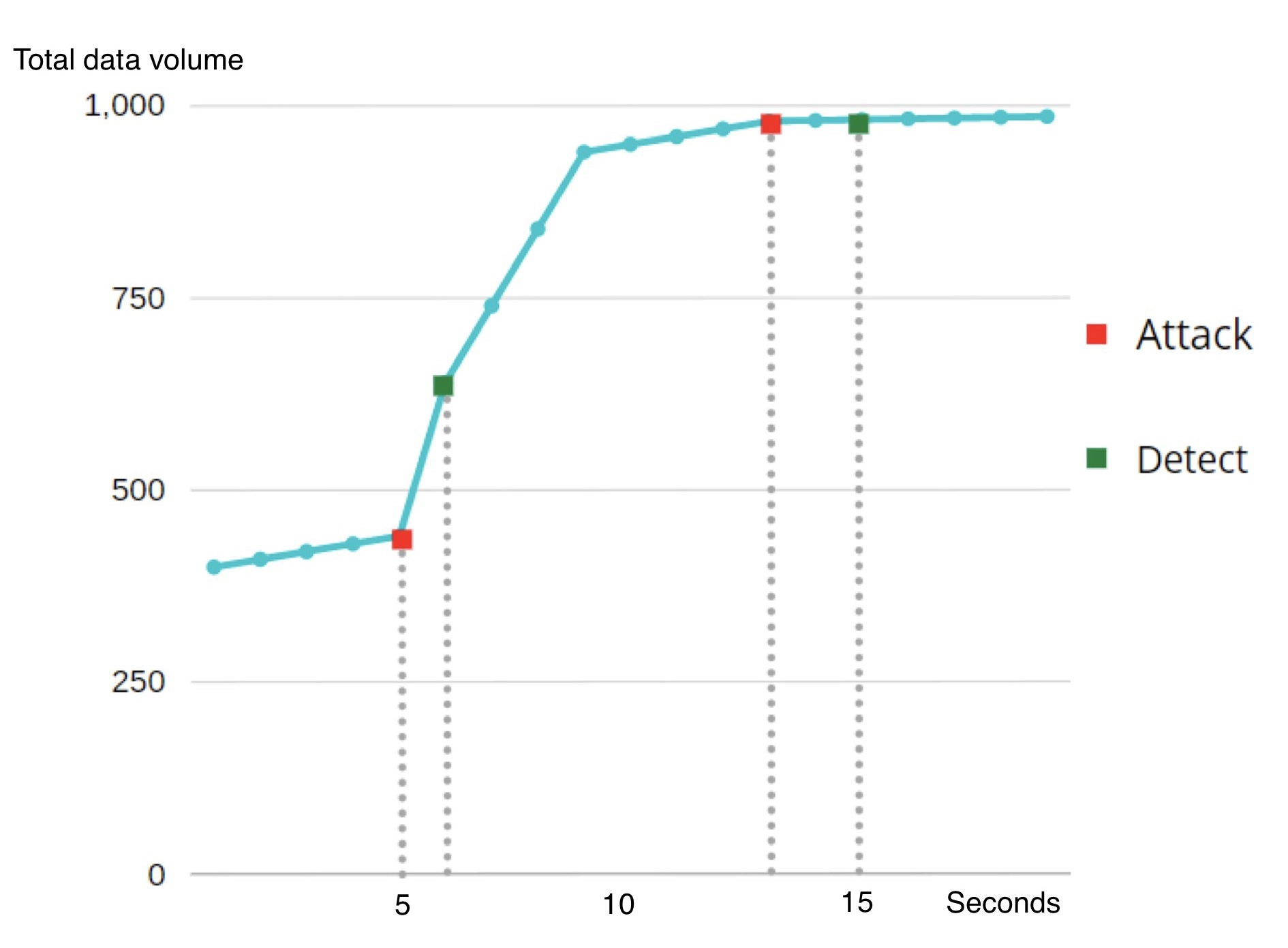}}
    \subfigure[]{\includegraphics[width=0.48\linewidth]{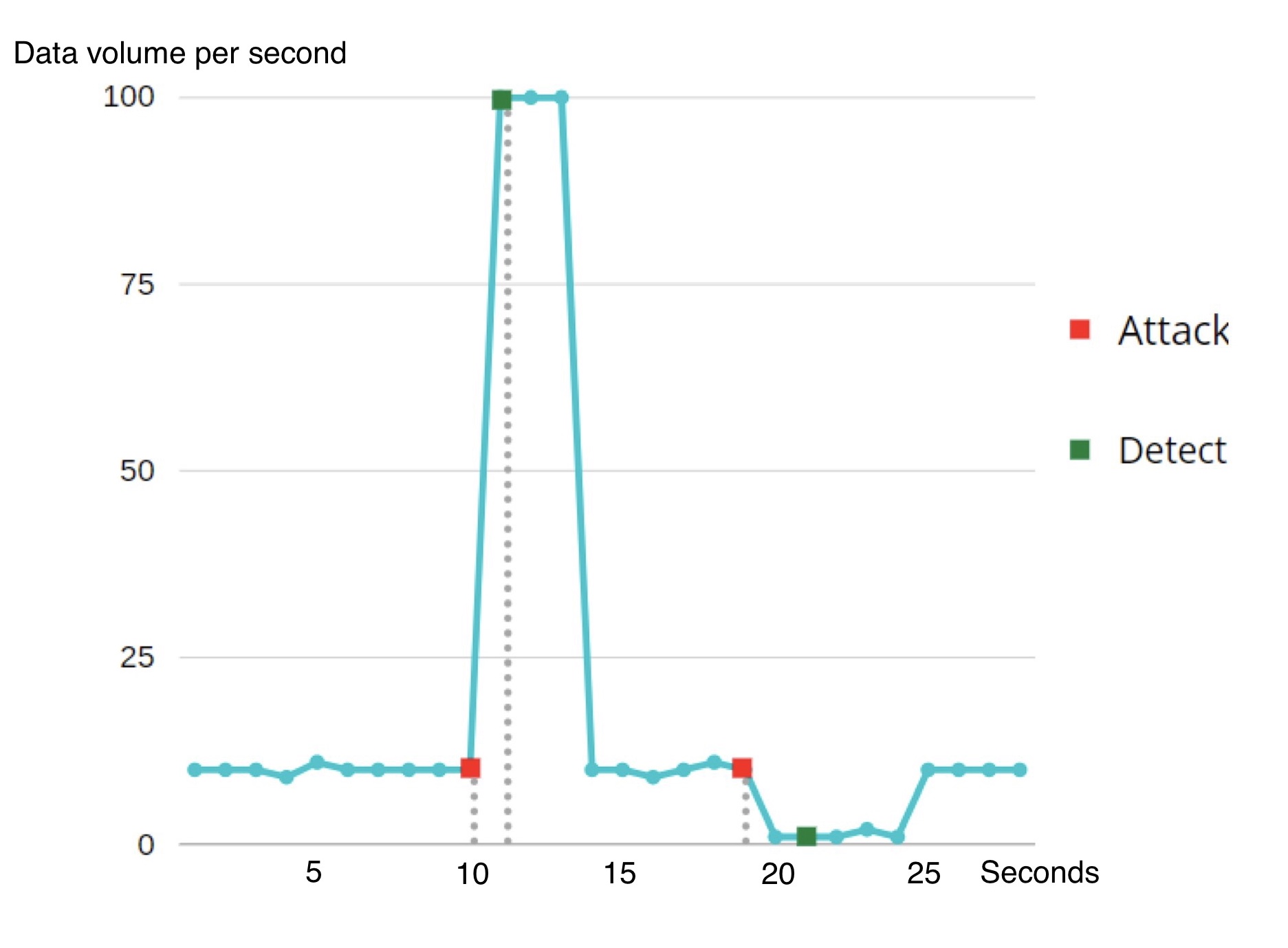}}
  \caption{(a) The data volume aggregated overtime; (b) change of data volume uploaded per second. The system will detect Flooding Attack in about 1 second; and Selective Forwarding Attack in about 2 seconds.}
    \label{fig:experiment_result}
    \vspace{0.2In}
\end{figure*}

\subsubsection{Implementing attack prevention and monitoring}
The IoT server is configured to accept only specified data formats when receiving data uploaded by nodes for the MQTT protocol. The data will be uploaded one by one in the format of "name: value" with the MQTT protocol to the virtual IoT server. The data that does not meet the requirements of the format will be filtered by the virtual IoT server, and the filtered data will be disassembled into field names and values, also the time stamp when uploaded to the server. The data will be stored in the specified table in the cloud database, and will be automatically marked with a serial number. This prevents Flooding, Jamming, and Exhaustion, which are attacks that transmit malicious data in large quantities in an attempt to exhaust the server's resources.

Also, when the IoT server detects a large amount of incorrectly formatted data being uploaded, it determines that an attacker is suspected of carrying out the attack, and then the server transmits the data to the cloud database, generates an alert record, and displays it in a web page.

When the IoT server forwards the data to the cloud database, it will perform data disassembly and data extraction according to our pre-defined SQL statements. If there is still data with empty or meaningless content at this point, it will be filtered out, so any malicious data that was not filtered out in the previous step will be filtered out in this step. This can also prevent Flooding attacks and Tampering attacks from contaminating the normal data stored in the database.

When a large amount of such data is filtered out, the amount of data received in the cloud database drops significantly. This will be detected as abnormal changes in the amount of data when the web back end queries for real-time data, thus an alert message will be generated.

If when the web back end queries the real-time data, it finds that there is no data update for a certain time, it is determined that there may be Selective Forwarding, Black Hole Attacks, Sinkhole Attacks and Misdirection Attacks, and an alert will be generated.

If when the web back end queries the real-time data, it finds that there is a large amount of data delayed for a while, it is judged that there may be a wormhole attack and alerted.

To prevent attackers from directly attacking the cloud database and cloud servers, we set up a white list for their connections, and only the IP address defined in the white list can transmit and read/write data. This also prevents attackers from transmitting a large number of packets to the cloud server disguised as nodes.

The database connection on the back end of the web page uses MyBatis, which prevents the user from entering direct use for database query statements in the web page and prevents SQL injection.

The GPS sensor uploads latitude and longitude information every 20 seconds in case of a physical attack, such as an attacker moving the node, or turning it off. If there is an apparent change in the coordinates of the GPS value display within a certain time. A warning to tampering attacks will be issued.

The web page is positioned using the Google Maps API, which marks the exact location of the sensor on the map based on the real-time coordinates obtained from the GPS in the database. Constant location uploading using GPS sensors is more accurate than direct positioning using GPRS, which has the potential to cause inaccurate data due to reasons such as base stations being too far away or delays, thus preventing the system from sending out timely warning messages.

% \begin{figure*}[t]
% 	\begin{minipage}[t]{0.48\linewidth}
% 	\centering
% 	\includegraphics[width=\linewidth]{figures/Total (2).PNG}.
% 	\caption{(a) The data volume aggregated overtime; (b) Change of data uploaded per second. The system will detect Flooding Attack in about 1 second; and Selective Forwarding Attack in about2 seconds.}
% 	\label{fig:fig3}
% 	\end{minipage}
% 	\begin{minipage}[t]{0.48\linewidth}
% 	\centering
% 	\includegraphics[width=\linewidth]{figures/Persecond (2).PNG}.
% 	\caption{Amount of data uploaded per second.}
% 	\label{fig:fig4}
% 	\end{minipage}
% \end{figure*}

% \begin{figure}
%     \centering
%     \includegraphics[width=0.98\linewidth]{figures/Total (2).PNG}
%     \caption{The data volume aggregated overtime.}
%     \label{fig:fig3}
% \end{figure}

% \begin{figure}
%     \centering
%     \includegraphics[width=0.98\linewidth]{figures/Persecond (2).PNG}.
% 	\caption{Change of data uploaded per second. The system will detect Flooding Attack in about 1 second; and Selective Forwarding Attack in about2 seconds.}
% 	\label{fig:fig4}
% \end{figure}

\section{Evaluation}
To verify the basic functions of this system, especially the attack prevention and monitoring part, we conducted some experiments. We simulated several common attacks to experiment the ability of the system against attacks. We also collected data on warning attacks and evaluated the timeliness of the warnings.

\subsection{Experiment setup}
To verify the system, we placed sensors in a real environment and collected data continuously for 1 month. We also regularly monitored the change of data volume in the web page and the result of data comparison.
Common attacks such as selective forwarding, black hole attacks, sinkhole attacks, flooding attacks and misdirection attacks can be classified as three types of data change attacks. They are large amounts of data type, data loss type and data delay type attacks. 

We simulate the large amounts of data type attack by uploading a large amount of data and empty packets from the development board. For the data loss type of attack, we simulate it by lowering and stopping the upload of data within a certain time. For the data delay type of attack, we simulate the data delay by modifying the timestamp of the collected data before uploading it from the development board. In addition to the above, we also simulate physical attacks and cloud database and server attacks.

To evaluate this system, except simulated abnormal upload data volume changes to simulate attacks in several experiments, we also compared the average time for the system to issue an alert with the time when an attack actually occurs.
 
\subsection{Experimental results}
When a large amount of data comes into the cloud server, it is rejected because the data is not in the right format so that it does not affect the data stored in the cloud database. For the data loss type of attack, when the page was refreshed, the back end looked for data and found no data update, so the system successfully detected the attack. For the data delay type of attack, The system found a large number of data with too large interval between receiving time and uploading time when querying data, so it made an early warning. After we simulated an abnormal change in GPS location, the physical attack was successfully being warned. For the cloud database and virtual server attacks, we used a new IP address to upload data to it, and it was rejected. This successfully protected the server from the fake nodes.

The curves in Figure~\ref{fig:experiment_result} show the aggregated data volume over time and the data volume changed per second when a large number of data type attacks and data loss type attacks occur.

In Figure~\ref{fig:experiment_result}, the warning is triggered when the uploading data volume rises by 10\% or drops by 4\% in the last 40 seconds compared to the data in the last 80-40 seconds. As can be seen from the Figure~\ref{fig:experiment_result}(b), when the attack starts, the system will detect Flooding Attacks in about 1 second and Selective Forwarding Attack in about 2 seconds. Also, since there is a 10\% and 4\% up and down range, it can effectively prevent misjudgment. Therefore, it can be considered that the warning is issued accurately and timely.

\section{Conclusion}
The system implemented in this paper is capable of monitoring and analyzing agricultural data, as well as preventing and monitoring attacks. The temperature and humidity data are collected by using sensors, the location data are collected by GPS module and they are transmitted to the virtual IoT server by GPRS module in MQTT protocol. The data is then filtered and processed to prevent attacks, and finally the data is forwarded to a cloud database for storage.

The system monitors real-time agricultural data and provides analysis services through a web page. It can detect the presence of attacks and issue warnings on the page by querying the data in the database and analyzing the data. The web page also provides queries and comparisons of historical data, analysis of changes in data volumes, and warning of extreme temperature and humidity.

This system was validated in experiments and the results showed that it is effective in preventing and monitoring common attacks such as Selective Forwarding, Black Hole Attacks, Sinkhole Attacks, Flooding Attacks, and Misdirection Attacks.

Further research could be done in the area of data protection, such as encrypting data or taking protective measures against data loss. Additional methods could also be used to make the monitoring of attacks more accurate so that the system can more clearly distinguish between each type of attack.

In the hardware transfer to the server part, multiple protocol types can be used simultaneously for uploading, not only using the MQTT protocol, but also using the TCP protocol for uploading, using data cross-referencing to both prevent and detect attacks.

% \newpage
\balance
\bibliographystyle{unsrt}
\bibliography{ref}

\end{document}